\documentclass[12pt,preprint]{aastex}
\usepackage{color}
\newcommand{\hi}{H\,I}
\newcommand{\sdssj}{J0952$+$0114}
\newcommand{\siii}{Si\,{\scriptsize II}}
\newcommand{\feii}{Fe\,{\scriptsize II}}
\newcommand{\znii}{Zn\,{\scriptsize II}}
\newcommand{\nv}{N\,{\scriptsize V}}
\newcommand{\civ}{C\,{\scriptsize IV}}
\newcommand{\ciii}{C\,{\scriptsize III]}}
\newcommand{\oi}{O\,{\scriptsize I}}
\newcommand{\mgi}{Mg\,{\scriptsize I}}
\newcommand{\crii}{Cr\,{\scriptsize II}}
\newcommand{\lla}{$L_{{\rm Ly}\alpha}\sim10^{45}$~erg~s$^{-1}$}
\newcommand{\nhi}{$\log N_{\rm H\,I}({\rm cm}^{-2})=21.8\pm0.2$}
\shorttitle{Strong Ly$\alpha$ Emission in a PDLA Trough}
\shortauthors{Jiang et al.}
\begin{document}
\title{Strong Ly$\alpha$ Emission in the Proximate Damped Ly$\alpha$ Absorption Trough toward the Quasar SDSS J095253.83$+$011422.0}
\author{Peng Jiang\altaffilmark{1,2,3,4}, Hongyan Zhou\altaffilmark{3,4}, Xiang Pan\altaffilmark{4,3}, Ning Jiang\altaffilmark{4},Xinwen Shu\altaffilmark{5}, Huiyuan Wang\altaffilmark{4}, Qiusheng Gu\altaffilmark{1,2}, Zhenzhen Li\altaffilmark{4,3}, Maochun Wu\altaffilmark{4}, Xiheng Shi\altaffilmark{3,4}, Tuo Ji\altaffilmark{3,4}, Qiguo Tian\altaffilmark{3,4}, and Shaohua Zhang\altaffilmark{3,4}}
\altaffiltext{1}{School of Astronomy and Space Science, Nanjing University, Nanjing, Jiangsu, 210093, China}
\altaffiltext{2}{Key Laboratory of Modern Astronomy and Astrophysics (Nanjing University), Ministry of Education, Nanjing, Jiangsu, 210093, China}
\altaffiltext{3}{Polar Research Institute of China, Jinqiao Rd. 451, Shanghai, 200136, China}
\altaffiltext{4}{Key Laboratory for Research in Galaxies and Cosmology, The University of Science and Technology of China, Chinese Academy of Sciences, Hefei, Anhui, 230026, China}
\altaffiltext{5}{Department of Physics, Anhui Normal University, Wuhu, Anhui, 241000, China}
\email{zhouhongyan@pric.org.cn}

\begin{abstract}
SDSS J095253.83$+$011422.0 (\sdssj) was reported by Hall et al. (2004) as an exotic quasar at $z_{em}=3.020$.
In contrast to prominent broad metal--line emissions with FWHM$\sim9000$~km~s$^{-1}$, only a narrow Ly$\alpha$
emission line is present with FWHM$\sim$1000~km~s$^{-1}$. The absence of broad Ly$\alpha$ emission line has been a mystery for more than a decade. In this paper,
we demonstrate that this is due to dark Proximate Damped Ly$\alpha$ Absorption (PDLA) at
$z_{abs}=3.010$ by identifying associated Lyman absorption line series from the damped Ly$\beta$ up to Ly9, 
as well as the Lyman limit absorption edge.
The PDLA cloud has a column density of \nhi~, a metallicity of [Zn/H]$>-1.0$, and a spatial extent exceeding the Narrow Emission Line Region (NELR) of the quasar. 
With a luminosity of \lla, the residual Ly$\alpha$ emission superposed on the PDLA trough is of two orders of magnitude stronger than previous reports.
This is best explained as re-radiated photons arising from the quasar outflowing gas at a scale larger than the NELR.
The PDLA here, acting like a natural coronagraph, provides us with a good insight into the illuminated gas in
the vicinity of the quasar, which are usually hard to resolve due to their small size and ``seeing fuzz" of bright quasars.
Notably, \sdssj~analogies might be easily omitted in the spectroscopic surveys of DLAs and PDLAs, as their damped Ly$\alpha$ troughs can be fully filled by additional strong Ly$\alpha$ emissions. 
Our preliminary survey shows that such systems are not very rare. They are potentially a unique sample for probing strong quasar feedback phenomenon in the early universe.
\end{abstract}


\section{Introduction}
Damped Ly$\alpha$ absorption systems (DLAs; reviewed in Wolfe et al. 2005) with high neutral hydrogen column density are usually considered
as the main gas reservoir for
star formation in the high--redshift universe (e.g. Wolfe et al. 1986; Nagamine et al. 2004; Prochaska et al. 2005). As the emission of background quasar is
completely absorbed in the DLA trough, it has been proposed that the Ly$\alpha$ emission from star formation in DLA host galaxy can be
observed with an assistant from the natural coronagraph. However, most of searches for the residual Ly$\alpha$ emission in broad DLA trough of quasar spectra
resulted in non--detections (e.g. Smith et al. 1989; Kulkarni et al. 2006; Cai et al. 2014). Individual detections were sparsely reported,
e.g. in the super--DLA with $N_{\rm \hi} \approx 10^{22.1}$~cm$^{-2}$ toward  Q1135$-$0010 (Noterdaeme et al. 2012a; Kulkarni et al. 2012),
in the strong DLA with $N_{\rm \hi} \approx 10^{21.6}$~cm$^{-2}$ toward PKS~0458$-$02 (M{\o}ller et al. 2004), 
and in the metal-rich DLA towards Q2222-0946 (Fynbo et al. 2010). Recently,
Noterdaeme et al. (2014) studied a large sample of extremely strong DLAs (ESDLA; $N_{\rm \hi} \geqslant 10^{21.7}$ cm$^{-2}$) observed
in BOSS of SDSS--III\footnote{SDSS: Sloan Digital Sky Survey; BOSS: Baryon Oscillation Spectroscopic Survey}
and found significant residual Ly$\alpha$ emission in the composite ESDLA spectrum. The ESDLAs are plausibly associated with
galaxies having high star formation rates.

Ly$\alpha$ ``blobs'' have been discovered occasionally at close redshifts to quasars with tiny impact parameters of $\lesssim$~5\arcsec~(Weidinger et al. 2005;
Zafar et al.2011). These Ly$\alpha$ photons are most probably radiated from ambient gas illuminated by their associated quasars (Weidinger et al. 2004).
On the other hand, DLAs in proximity to quasars with $z_{abs} \approx z_{em}$ (PDLAs; Prochaska et al. 2008) appear to
preferentially exhibit residual Ly$\alpha$ emissions in their absorption troughs (e.g. M{\o}ller \& Warren 1993; M{\o}ller et al. 1998;
Ellison et al. 2002; Hennawi et al. 2009; Finley et al. 2013; Fathivavsari et al. 2015). Similar to Ly$\alpha$ ``blobs'',
fluorescent recombination radiation from ambient gas photoionized by the quasar,
rather than star formation activity in quasar host galaxy, is likely the main contributor to the residual Ly$\alpha$ emission found in PDLAs.
This is the case for the strong Ly$\alpha$ emission ($L_{{\rm Ly}\alpha}=3.9\times10^{43}$~erg~s$^{-1}$) superposed on the trough of
the PDLA toward the quasar SDSS J124020.91$+$145535.6 (hereafter J1240$+$1455; $z=3$; Hennawi et al. 2009). The residual Ly$\alpha$ emission
is resolved by long--slit spectroscopy as a Ly$\alpha$ blob with spatial extent exceeding 5\arcsec~around the quasar (e.g. Christensen et al. 2006;
Hennawi et al. 2009). Finley et al. (2013) identified 26 PDLAs showing residual narrow Ly$\alpha$ emissions with strengths similar to that of
the Ly$\alpha$ emission observed in J1240$+$1455 from DR9 and DR10.

In the course of a survey for the quasar intermediate--width emission lines (IELs; Li et al. 2015) in SDSS DR12\footnote{http://www.sdss.org/dr12/},
we found a high--confidence candidate of Ly$\alpha$ IEL in the spectrum of the quasar SDSS J095253.83$+$011422.0
(hereafter J0952$+$0114) at $z=3.020$. The observed Ly$\alpha$ line is
extremely narrow (FWHM$\sim$1000 km s$^{-1}$) in comparison with the broad metal emission lines (FWHM$\sim$9000 km s$^{-1}$). The absence
of broad Ly$\alpha$ emission line in J0952$+$0114 was first reported by Hall et al. (2004), using the SDSS--I/II spectrum. The authors suggested
that the broad line region might be dominated by very dense gas (10$^{15}$~cm$^{-3}$) in which broad Ly$\alpha$ emission is suppressed, but unusual
ionization parameters and peculiar configuration of emitting clouds are required to reproduce the emission line spectrum.
An alternative explanation that the broad Ly$\alpha$ emission is removed by smooth absorption arising from \nv~and Ly$\alpha$, was discussed
as well, but regarded as unlikely there (Hall et al. 2004).

The BOSS spectra of J0952$+$0114, covering a wider wavelength range than SDSS spectrum, provide new insight into
this exotic quasar. The damped Ly$\beta$ absorption, the high--order Lyman--series absorptions, and the Lyman limit absorption edge at
$z_{abs}\approx$~$z_{em}$ are firmly detected at the blue end of its two BOSS spectra. These detections demonstrate that a high column density PDLA
intercepts the sightline to the quasar. The damped Ly$\alpha$ absorption trough removes the intrinsic broad
Ly$\alpha$ emission of the quasar completely. The observed narrow Ly$\alpha$ emission line should arise from the ambient gas in an
extended area, which is larger than the cross--section of the PDLA cloud. The geometry of this system (i.e. the background quasar, the
PDLA cloud and the emitting gas) is similar to that of the system seen toward the quasar J1240$+$1455 (Hennawi et al. 2009).
However, the luminosity of residual Ly$\alpha$ emission observed in J0952$+$0114 is much larger than that in SDSS J1240$+$1455 by nearly two orders
of magnitude, plausibly indicating that they have different origins. 

This paper is organized as follows. In \S2, we summarise the observations; detailed analysis of the absorption and emission spectrum are presented in
\S3; in \S4, we discuss the origin of the luminous Ly$\alpha$ emission in J0952$+$0114 as well as applications and implications;
a summary is then made in \S5.
Throughout this paper, $H_0$ is taken to be 70.3 km s$^{-1}$ Mpc$^{-1}$, $\Omega_m=0.27$, and $\Omega_\Lambda=0.73$.

\section{Observations}
J0952$+$0114 is a quasar at $z=3.020$ identified by SDSS. Two spectra were first acquired with the SDSS spectrograph
(York et al. 2000) fed with fibers in diameters of 3\arcsec~on 2000 March 05 and 2000 December 30. The spectra spread a wavelength range
of $\lambda\sim$3,800--9,200~\AA. Two additional spectra were taken using the BOSS spectrograph (Dawson et al. 2013), which are fed with
2\arcsec~fibers and provide a more extended wavelength coverage of $\lambda\sim$3,600--10,400~\AA, on 2010 December 09
and 2011 March 25. All spectra share a similar spectral resolution of $R\sim$2,000. We carefully compared the SDSS spectra with
the BOSS spectra in continuum, emission lines and absorption lines. No significant variation in either of these spectral
features spanning $\sim$ 2 years in quasar's rest-frame is detected. Figure 1 presents the combined BOSS spectrum of
J0952$+$0114.\footnote{The dust map of Schlafly \& Finkbeiner (2011) is implemented to correct the Galactic extinction.} The Ly$\alpha$ emission line
is significantly narrower than the broad metal
emission lines (see inserted panel in Figure 1). Moreover, an associated \civ~absorption system with plenty of absorption lines is identified at z$_{abs}=$3.010.

\section{Data Analysis and Results}

\subsection{Identification of PDLA}
The BOSS spectrum of J0952$+$0114 ranges from 895~\AA~to 2550~\AA~in quasar's rest-frame. It enables us to search for the neutral hydrogen
Lyman--series absorptions corresponding to the associated \civ~absorber (Figure 2a). Firstly, the damped absorption trough of
Ly$\beta$ is credibly identified, which is much broader than any other obvious absorption features (possibly intervening Ly$\alpha$ absorption lines) 
blanketing the spectrum at $\lambda<$~1216~\AA~
. The Lyman--series absorptions in higher orders, from Ly$\gamma$ to even Ly9, can be identified despite contamination from the Ly$\alpha$ forest.
In the very blue end of the spectrum, the flux at rest wavelengths $\lambda<$~912~\AA~is
deeply suppressed, suggesting an optically thick neutral hydrogen cloud at the Lyman limit. All the facts, especially the presence of the
damped Ly$\beta$ absorption, reveal that the associated absorber has a high column density of neutral hydrogen.

In order to normalize the observed spectrum,
we model the unabsorbed spectrum of J0952$+$0114 with the composite $HST$ quasar spectrum (Zheng et al. 1997) as an initial guess.
The quasar continua fit well with each other, but the observed emission line of Ly$\beta+$O\,{\scriptsize{VI}} is stronger than that in the
composite spectrum (Figure 2a). In the composite $HST$ spectrum, the continuum in ultraviolet (UV) can be modelled as a broken power law, with 
the spectral index of $\alpha=-2.2$ at 350~\AA~$\leqslant \lambda \leqslant$ 1050~\AA~ ($f_{\nu}\propto{\nu}^{\alpha}$) and $\alpha=-0.99$ at
1050~\AA $\leqslant \lambda \leqslant$ 2200~\AA~is  (Zheng et al. 1997). We model the UV continuum of J0952$+$0114 using the same broken power law and
model the Ly$\beta+$O\,{\scriptsize{VI}} emission line with one broad Gaussian. The full model of the unabsorbed spectrum is presented
in Figure 2a and is applied for normalization. 

We measure the column density of neutral hydrogen using Voigt profile fitting to the Ly$\beta$ trough (Figure 2b).
The best--fit model well reproduces the damped wings of Ly$\beta$,
while there is very weak residual flux in the line center. We infer that the residual flux is the Ly$\beta$ emission
associated with the strong Ly$\alpha$ emission of interest.
The Voigt profile fitting results in a column density of $\log N_{\rm \hi}({\rm cm}^{-2})=21.8\pm0.1$.
Considering the systematic error due to uncertainty of normalization, we conservatively increase the error by 0.1 dex and adopt
$\log N_{\rm \hi}({\rm cm}^{-2})=21.8\pm0.2$. We have demonstrated that the associated absorber toward J0952$+$0114 is a very strong PDLA.
Any Ly$\alpha$ emission from the regions covered by the PDLA cloud in the line of sight should be
blocked completely. The observed Ly$\alpha$ emission in J0952$+$0114 necessarily arises from extended regions around the quasar nucleus
(or in the host galaxy), which is beyond the coverage of the PDLA cloud.

Jenkins (1990) proposed a technique to deduce the velocity dispersion of absorbers from the convergence of the Lyman--series lines at
wavelengths just above the Lyman limit. The Lyman line convergence technique is very useful for low--resolution spectra
(Hurwitz \& Bowyer 1995). However, the Lyman edge cannot be confidently identified in the spectrum of \sdssj, owing to heavy contamination
from Ly$\alpha$ forest absorptions and the relatively low signal--to--noise ratio of data at this spectral region. Instead, we estimate the Doppler parameter 
$b$ by fitting the depths of high--order Lyman--series lines. Having such a high neutral hydrogen column density of
$\log N_{\rm \hi}({\rm cm}^{-2}) \sim 21.8$, the Lyman--series absorption lines, in orders up to Ly9, should be saturated. The non--zero
fluxes observed at positions of the line centers (Figure 2c) indicate that the lines are not resolved. If the instrumental resolution is given,
we can express the line depths as functions of $b$ solely. Adopting the derived $N_{\rm \hi}$ and the spectral dispersion function provided
along with BOSS spectra, we vary $b$ to fit the observed depths of Lyman--series absorption lines.
A best--fit model with $b=35$~km~s$^{-1}$ is derived, and a minimum width of $b=55$~km~s$^{-1}$ is found for the models having zero fluxes at the centers
of Lyman series up to Ly9, which appears to disagree with the observation.
Since the high--order Lyman--series lines are contaminated by Ly$\alpha$ absorption forest seriously, it is difficult to gauge the
uncertainty of $b$. In a conservative manner, the fairly loose upper limit of $b<55$~km~s$^{-1}$ is adopted in the following analysis.

\subsection{\ciii~and~\civ~Emission Lines}
The \ciii~and~\civ~lines are the strongest broad emission lines in the BOSS spectrum of J0952$+$0114 (Figure 3a).
The narrow \civ~and~\ciii~emission lines are strong, if compared with the composite $HST$ quasar spectrum.
The line profiles of \civ~and~\ciii~are obtained by subtracting a power--law continuum from the observed spectrum, 
presented in velocity space in Figure 3b and 3c, respectively. 
We fit the line profile of \ciii~with four Gaussians (Figure 3b). Three Gaussians are used to model the complex profile of the broad component
and the other one for the narrow component. The narrow \ciii~emission line is so prominent that it is fairly well
constrained, yielding FWHM$=532\pm37$~km~s$^{-1}$. Since the associated \civ~absorption lines are extremely narrow, the line profile of
\civ~emission is mostly preserved. We directly mask the absorptions tough and fit the remaining profile using four Gaussians as well (Figure 3c).
A strong narrow \civ~emission line, with FWHM$=631\pm48$~km~s$^{-1}$, is firmly detected. In a brief summary, both of the
\ciii~and~\civ~emission lines of J0952$+$0114 are decomposed into a complex broad component arising from the broad emission line region (BELR)
and a narrow Gaussian component arising from the narrow emission line region (NELR).

Furthermore, the coverage of the PDLA to NELR, the outermost region of the quasar J0952$+$0114's nucleus, is investigated, by analysing the line ratios of the isolated \civ~absorptions
in different cases. Assuming the absorber fully covers the nucleus,
we divide the observed spectrum by the full model of quasar emission in \civ~region to normalize the \civ~absorption lines.
The normalized spectrum is presented in Figure 4a. The rest--frame equivalent widths (EWs) are EW(\civ$\lambda$1548)$=0.69\pm0.03$\AA~and
EW(\civ$\lambda$1550)$=0.60\pm0.03$\AA, respectively. If the absorber does not cover the NELR, the narrow \civ~emission line
should be subtracted before normalization. This yields EW(\civ$\lambda$1548)$=0.71\pm0.03$\AA~and EW(\civ$\lambda$1550)$=0.88\pm0.03$\AA~
(Figure 4b). The oscillator strength of \civ$\lambda$1548 is twice as large as that of \civ$\lambda$1550, thus their equivalent width ratio
should range from 2 (both of the absorption lines are in linear part of Curve of Growth) to nearly 1 (doublets are seriously saturated). In theory, the line ratio of \civ$\lambda$1548 and \civ$\lambda$1550 must be no less than one. The second scenario
, providing an unpractical line ratio, is therefore ruled out. The analysis indicates that the quasar nucleus of J0952$+$0114 is fully covered
by the associated absorber (i.e. the PDLA).

\subsection{The Residual Ly$\alpha$ Emission}
In our scenario, the spectrum of J0952$+$0114~in the Ly$\alpha$ band is the sum of the absorbed quasar emission, by the PDLA, and the residual
Ly$\alpha$ emission superposed on the absorption trough. The column density $N_{\rm \hi}$ of the PDLA has been obtained by fitting the associated
Ly$\beta$ damped trough. To model the absorbed quasar emission, we have to reconstruct the intrinsic quasar emission, which consists of a continuum, and intrinsic Ly$\alpha$ and \nv~emission lines. 
The quasar continuum can be well modelled by a power law in a wide wavelength range (Figure 5a). However, the flux
on both sides of the Ly$\alpha$ emission is significantly lower than the model continuum. The suppressed flux can be naturally explained as
the damped wings of the associated Ly$\alpha$ absorption. Assuming the intrinsic Ly$\alpha$ and \nv~emission lines share the line profile
of \civ~emission, we utilize the best--fit profile of \civ~line to model them. The intrinsic
quasar emission is then absorbed by the PDLA with a fixed column density of $\log N_{\rm \hi}({\rm cm}^{-2})=21.8$. We adjust the strengths of the
model emission lines to fit the partially observed damped wings of PDLA. The residual Ly$\alpha$ emission is finally derived by subtracting the
best--fit model of the absorbed quasar spectrum from the observed spectrum (Figure 5b).

The blue wing of residual Ly$\alpha$ emission is destroyed by the Ly$\alpha$ forest absorptions. In order to repair its profile, we model the
Ly$\alpha$ emission with two Gaussians by requiring the model to fit the red wing and predict a blue wing to match the
imperfect envelope. The width of the repaired Ly$\alpha$ line is FWHM$=1106\pm75$ km s$^{-1}$. The total flux of the residual Ly$\alpha$ emission is
$f=1.64\pm0.08\times10^{-14}$~erg~s$^{-1}$~cm$^{-2}$, yielding a luminosity of $L_{{\rm Ly}\alpha}=1.36\pm0.06\times10^{45}$~erg~s$^{-1}$.
Directly integrated residual line flux is also calculated, which is $f=1.23\times10^{-14}$~erg~s$^{-1}$~cm$^{-2}$,
 and the corresponding luminosity is
$L_{{\rm Ly}\alpha}=1.02\times10^{45}$~erg~s$^{-1}$.

\subsection{Associated Absorption Lines of Heavy Elements}
The associated absorber at $z_{abs}=3.010$ imprints plenty of metal absorption lines in the spectrum of \sdssj.
The absorption lines in the featureless continuum are normalized straightforwardly: dividing the observed spectrum by a power law
fitted to the local continuum. For the absorption lines superposed on emission lines, we first fit the emission profiles with
one to four Gaussians and then divide the observed spectrum by the best--fit model.

\znii$\lambda\lambda$2026,2062 absorption lines are
detected in the BOSS spectrum. However, these two absorption lines are possibly blended with \mgi$\lambda$2026 and \crii$\lambda$2062 lines,
respectively. A search for the stronger \crii$\lambda$2056 line associated with the \crii$\lambda$2062~results in a non-detection, suggesting
that \znii$\lambda$2062 absorption is almost isolated. Therefore, the detections of \znii$\lambda$2062 line and the corresponding stronger
\znii$\lambda$2026 line are solid.
High--ionized absorption lines and lines from excited fine-structure levels are useful diagnostics for the physical parameters of quasar
absorbers (Ellison et al. 2010). \nv$\lambda\lambda$1238,1242 absorption lines are clearly detected in the spectrum.
Three lines of \siii$^*$ (i.e. \siii$^*$$\lambda$1264, \siii$^*$$\lambda$1309 and
\siii$^*$$\lambda$1533) are tentatively detected, but \siii$^*$$\lambda$1264 and
\siii$^*$$\lambda$1533 lines are then identified to be an intervening \feii$\lambda$2382 absorption line at $z=1.13$ and a \civ$\lambda$1548 absorption line at
$z=2.97$. The isolated absorption line of \siii$^*$$\lambda$1309 looks strong but its profile is too much broader than that of other
metal lines. We conclude that all the detections of \siii$^*$~transitions are false positives.
The \oi$^*$$\lambda$1304 absorption line is detected, but it is seriously blended with \siii$\lambda$1304 and
\oi$^{**}$$\lambda$1306 lines (Figure 6). None of the absorption lines from excited states of Fe$^+$ is detected in the BOSS spectrum of \sdssj.
We illustrate several absorption lines in Figure 6.

EWs of the absorption lines are measured by integrating the normalized spectrum over a velocity range of
[-400,400]~km~s$^{-1}$. The results are summarized in Table 1. In \S3.1, we have obtained the upper limit of the Doppler parameter
(i.e. $b<55$ km~s$^{-1}$). In Figure 7, we present four curves of growth (COG; Jenkins 1986) with Doppler parameters of $b=55$ km~s$^{-1}$,
$b=35$ km~s$^{-1}$, $b=25$ km~s$^{-1}$ and $b=15$ km~s$^{-1}$, respectively.

We can see that the shape of COG changes dramatically with $b$. COG with smaller Doppler parameters are having narrower linear part. 
Thus reducing the value of $b$ gradually from 55 km~s$^{-1}$ to 15 km~s$^{-1}$ will cause most of the detected absorption lines 
moves into of the logarithmic part, in which column density is highly sensitive to both $b$ and EWs. A reliable column densities measurement using
the metal absorption lines detected in J0952$+$0114 is therefore unable to be achieved, given that the Doppler
parameter cannot be determined in advance (Prochaska 2006). We locate the metal absorption lines on the COG of $b=55$ km~s$^{-1}$.
The column densities measured on it are therefore conservative lower limits. In this manner, we estimate the column density of Zinc,
yielding $N_{{\rm Zn}^+}>10^{13.4}$~cm$^{-2}$. The relative abundance
to the solar value is [Zn/H]$>-1.0$, i.e. in the range of metal--strong DLAs (e.g. Kaplan et al. 2010; Wang et al. 2012).
It is worth following up the PDLA toward J0952$+$0114 using high--resolution spectroscopy to measure the column densities
more precisely. 

\section{Discussion}
\subsection{PDLA Cloud}
In \sdssj, the requirement of proximity to a hard
ionizing source for the presence of \nv~transitions, which are rarely seen in intervening DLAs, constrains
the distance between the quasar and absorber of $d<1$ Mpc (Fox et al. 2009; Hennawi et al. 2009). Ellison et al. (2010) studied seven
PDLA systems with high--resolution spectroscopy, and estimated a lower limit on the distance $d>150$ kpc via the UV pumping rate estimated by
the strengths of \siii$^*$ absorptions. Thus, those PDLAs are unlikely to be associated with their quasar hosts but foreground galaxies in
proximity to the quasar. Recently, Fathivavsari et al. (2015) reported an intrinsic DLA toward the quasar SDSS J082303.22$+$052907.6.
The distance of the absorbing cloud to the quasar nucleus is $d<1$ kpc, and extended Ly$\alpha$ emission was observed with the assistance from the
coronagraphic cloud. However, the origin of the PDLA toward \sdssj~can be hardly determined in this way, as the \siii$^*$ transitions are
not detected in the spectrum.

The coverage analysis of the \civ~absorption lines has revealed that the PDLA cloud is in a scale larger than the NELR of \sdssj.
If the PDLA absorber is the inter--stellar medium (ISM) in the quasar host galaxy or in a foreground galaxy, its volume density might be
similar to regular DLAs, in a range of $n_{\rm H}\sim$1--100~cm$^{-3}$ (Prochaska \& Hennawi 2009). Assuming a homogeneous density in the cloud,
we estimate its length scale as $r=N_{\rm H}/n_{\rm H}\approx$20~pc--2~kpc. It is consistent with, more extended in a way, the scales of NELR observed
in the local active galactic nuclei (AGNs; e.g. Kraemer et al. 1994; Hutchings et al. 1998; Peterson et al. 2013).
If the PDLA cloud has a higher density, it would be necessarily closer to the quasar nucleus, more likely in the quasar host galaxy,
to keep a relatively high ionization parameter $U$ to produce \nv~absorption lines. Ji et al. (2015) identified an intrinsic absorber,
via its He\,{\scriptsize I}$^*$ and \feii$^*$ absorption lines, toward the quasar SDSS J080248.18$+$551328.9 and derived a density
of $n_{\rm H}\sim10^5$~cm$^{-3}$, a distance of 100--250~pc to the nucleus. The PDLA of interest is unlikely to have such a high density,
or the corresponding size will be too small to cover the quasar NELR in that case.

\subsection{Origin of the Luminous Ly$\alpha$ Emission}
The quasar \sdssj~was initially selected as an IEL quasar candidate. However, the Ly$\alpha$ IEL scenario has to be ruled out. As the
NELR, i.e. the outermost region of quasar nucleus, is already covered by the PDLA cloud, the IEL region that presumably exists between
the BELR and NELR (e.g. Brotherton et al. 1994), must be blocked in the Ly$\alpha$ band.

Outflows over NELR scales have been observed widely, traced by the optical emission lines of their photoionized
gas, in nearby AGNs (e.g. Fu \& Stockton 2009; Liu et al. 2013; Harrison et al. 2014). The emission lines arising
from those outflows generally have velocity widths of $\sim$800 km s$^{-1}$, which are above the escape velocities from the host galaxies.
Having a similar line width, the Ly$\alpha$ emission of \sdssj~is very likely the fluorescent recombination radiation from outflows driven
by quasar. We derive a covering factor of the emitting gas as $cf=L_{Ly\alpha}/(0.6h\nu)/Q=9\%$,
where the ionizing photon rate $Q\sim1.1\times10^{57}$ s$^{-1}$ is
estimated by scaling the observed spectrum with the MF87 SED (Spectral Energy Distribution; Mathews \& Ferland 1987) and the case B recombination
is assumed. This is over an order of magnitude larger than that of the Ly$\alpha$ blobs observed around
quasars ($cf\approx$0.5\%; e.g. Heckman et al. 1991; Christensen et al. 2006). The large covering factor
also supports the outflow scenario rather than quiescent ISM or infalling gas in/onto the quasar host galaxy
(Rees 1988; Haiman \& Rees 2001; Hennawi et al.2009).

The spatial extent of outflows is critical to the evaluation of whether a quasar is able to input its energy effectively into ISM of its host galaxy.
Only galaxy--scale outflows are efficient quasar feedbacks, which can quench the star formation widely in quasar hosts
and therefore regulate evolution of galaxies (e.g. Silk \& Rees 1998; Di Matteo et al. 2005). Meanwhile, the discoveries of
a significant population of massive evolved galaxies at $z>1.5$ (e.g. McCarthy et al. 2004; Saracco et al. 2005),
require that the quasar feedback must have impinged in the early universe. Observations of galaxy--scale outflows are desired to
study the quasar feedback at high redshifts. However, it is challenging to identify the extended outflows from the
``seeing fuzz" of bright quasars even if they are large enough to be spatially resolved. Currently, most of high--redshift galaxy--scale
outflows serving as evidences for quasar feedback in the early universe are actually observed in radio galaxies
and ultraluminous infrared galaxies (e.g. Nesvadba et al. 2008; Alexander et al. 2010; Harrison et al. 2012), where the central AGNs are
obscured and/or intrinsically dim. With the assistance of a PDLA blocking the emission at Ly$\alpha$ wavelengths from the background quasars,
the galaxy--scale outflows emitting substantial Ly$\alpha$ photons can be easily resolved, using
narrow--band imaging and long--slit/integral field spectroscopy, for high--redshift quasars in bright phase. 
The successful application of this technique in resolving the weak Ly$\alpha$ ``blob'' detected in the PDLA tough of the quasar Q0151+048 (Zafar et al. 2011) 
gives us further confidence in revealing outflows in \sdssj~analogues.

We then carefully re--examine the SDSS spectra with 3\arcsec~fiber and the BOSS spectra with 2\arcsec~fiber, of \sdssj, to seek for the possible
increment of Ly$\alpha$ flux observed in a larger aperture. There is no significant variation in the Ly$\alpha$ line, suggesting that the Ly$\alpha$
emission must be concentrated in 2\arcsec~on sky, i.e. a proper size of $\lesssim8$~kpc at $z=3.020$. In \sdssj, the luminous Ly$\alpha$ emission
contributes $\sim$8\% of the total flux in SDSS $g$ band. In Figure 8, we subtract its SDSS image by the point spread
function (PSF) model derived from the nearby stars in the field, which can be fitted using a 2D Gaussian with FWHM$=1.3$\arcsec.
No significant residual flux is detected, and therefore the luminous Ly$\alpha$ emission of interest is not spatially resolved in the SDSS image.

\subsection{Applications and Implications}
The approach of using DLAs as a natural coronagraph to observe the Ly$\alpha$ emission associated with star formation in their host galaxies
has been studied extensively (e.g. Kulkarni et al. 2006; Fynbo et al. 2010; Noterdaeme et al. 2014). The star formation rate (SFR) observed
in the high--redshift Lyman break galaxies (LBGs; Shapley et al. 2003; Erb et al. 2006) and Ly$\alpha$ emitters (LAEs; Gronwall et al. 2007)
are $\sim30M_{\sun}$~yr$^{-1}$, while the SFR in quasar host galaxies is $\sim9M_{\sun}$~yr$^{-1}$ (e.g. Ho 2005; Cai et al. 2014). Using the Kennicutt (1998)
calibration ${\rm SFR}(M_{\sun}\,{\rm yr}^{-1})=L(\rm H\alpha)/1.26\times 10^{41}$ erg s$^{-1}$ and assuming an intensity ratio of
${\rm Ly\alpha}/{\rm H\alpha}=8.3$ for the case B recombination, we estimate a Ly$\alpha$ luminosity of $\sim5\times10^{43}$~erg~s$^{-1}$ for
the strong star formation activities. This is much lower than the luminosity of the residual Ly$\alpha$ emission observed in \sdssj.
Conclusively, the residual Ly$\alpha$ emission associated
with star formation is fairly weak, in comparison with quasar emission, and thus the DLA troughs can be seen clearly in this case.

The identification of DLA trough becomes difficult, if the residual Ly$\alpha$ flux is large enough to fill it almost entirely. It is notable that
\sdssj~was omitted in all the spectroscopic surveys of DLAs and PDLAs in SDSS (e.g. Prochaska et al. 2008; Noterdaeme et al. 2012b; Finley et al. 2013).
More proper criteria for selecting \sdssj~analogies would be: quasars having significantly narrower 
Ly$\alpha$ emissions than their broad \civ~lines, jointly with associated damped absorptions from
Ly$\beta$ (if its absorption trough is preserved) or associated strong metal absorption lines.
Our preliminary survey shows that such systems are not very rare in SDSS quasars. The strong Ly$\alpha$ emissions in the candidate systems could be
IELs or outflows driven by quasar, while other explanations are needed for individuals. Follow--up observations with long--slit,
integral field and high--resolution echelle spectrograph as well as narrow--band imaging will be useful for exploring their natures.

PDLAs, acting like natural coronagraphs, provide us with a unique insight into the gaseous content in the vicinity of quasars. Meanwhile,
other types of natural coronagraphs exist and are enormously useful for investigations on quasars. Recently, Li et al. (2015) detected a mostly
pure Ly$\alpha$ IEL in the quasar OI~287 where the dusty torus, as a natural coronagraph, coincidently blocks the BELR but leaves the IEL
region observable. The unambiguous detection demonstrated the existence of quasar IELs, which has been in debate for decades
(e.g., Wills et al. 1993; Brotherton et al. 1994; Mason et al. 1996). Strong \nv~broad absorption line (BAL; e.g., Weymann et al. 1991;
Voit et al. 1993; Zhang et al. 2010) absorbers are elegant coronagraphs in scales of BELRs (Zhang et al. 2015), which are potential tools for
exploring the sub--pc structures in quasars.

\section{Summary}
We revisit the unusual Ly$\alpha$ emission of \sdssj~(Hall et al. 2004), motivated initially by searching for the quasar IELs (Li et al. 2015). 
A high column density ($\log N_{\rm \hi}({\rm cm}^{-2})=21.8\pm0.2$) neutral hydrogen cloud
in proximity to the quasar is identified via the detections of an associated damped Ly$\beta$ absorption trough, high--order
Lyman--series absorptions, and the Lyman limit absorption edge in the BOSS spectrum of \sdssj. Meanwhile, the line ratio of the associated \civ~absorptions
suggests that both of BELR and NELR of the quasar are covered by the thick cloud. These new clues merge into a scenario that
the absence of broad Ly$\alpha$ line is due to strong PDLA absorption and the residual Ly$\alpha$ emission superposed on the trough is
fluorescent recombination radiation from ambient gas outside of NELR. With a luminosity of $L_{{\rm Ly}\alpha}\sim10^{45}$~erg~s$^{-1}$,
the residual Ly$\alpha$ emission is the largest ever seen in either intervening or proximate DLA troughs. The coverage analysis firstly
rules out an IEL origin of the observed Ly$\alpha$ emission. Its relatively broad line width and high luminosity largely exclude the
possibility that it is associated with star formation activities, but suggest an origin from outflows driven by quasar. The proper size
of the outflow, inferred from SDSS spectra and images, is $\lesssim8$~kpc.

PDLAs, acting like natural coronagraphs, provides us with a good probe into the illuminated gas in the vicinity of quasars, which
is hardly resolved due to its small size or ``seeing fuzz" of bright quasars.
It is worthwhile to search for analogies of \sdssj~in large sky area spectroscopic surveys, such as SDSS and LAMOST (Zhao et al. 2012).
A significantly narrower Ly$\alpha$ emission line than broad \civ~line and an associated damped absorption from Ly$\beta$ are proper
criteria for selecting \sdssj~analogies. Our preliminary survey shows that such systems are not very rare.
Follow--up observations with long--slit, integral field and high--resolution echelle spectrographs as well as narrow--band imagings are useful for
exploring the nature of the selected systems. Some of the Ly$\alpha$ emitting gas might be massive galaxy--scale outflows around optically bright
high--redshift quasars, which will be direct evidences for efficient quasar feedback in the early universe but rarely seen nowadays.

\acknowledgements
The authors appreciate the enlightening suggestions from the anonymous referee, which helped to improve the quality of this paper.
This work is supported by National Basic Research
Program of China (973 Program, grant No. 2015CB857005, grant No. 2013CB834905), the NSFC grant (grant No. 11233002, 11203022, 11421303, 11473025, 11033007),  the SOC program (CHINARE2015-02-03).

Funding for SDSS-III has been provided by the Alfred P. Sloan Foundation,
the Participating Institutions, the National Science Foundation, and the
U.S. Department of Energy Office of Science. The SDSS-III web site is
http://www.sdss3.org/.

SDSS-III is managed by the Astrophysical Research Consortium for the Participating
Institutions of the SDSS-III Collaboration including the University of Arizona,
the Brazilian Participation Group, Brookhaven National Laboratory, Carnegie Mellon University,
University of Florida, the French Participation Group, the German Participation Group,
Harvard University, the Instituto de Astrofisica de Canarias, the Michigan
State/Notre Dame/JINA Participation Group, Johns Hopkins University, Lawrence Berkeley
National Laboratory, Max Planck Institute for Astrophysics, Max Planck Institute for
Extraterrestrial Physics, New Mexico State University, New York University, Ohio State
University, Pennsylvania State University, University of Portsmouth, Princeton University,
the Spanish Participation Group, University of Tokyo, University of Utah, Vanderbilt
University, University of Virginia, University of Washington, and Yale University.

\begin{figure}
\epsscale{1.0}
\plotone{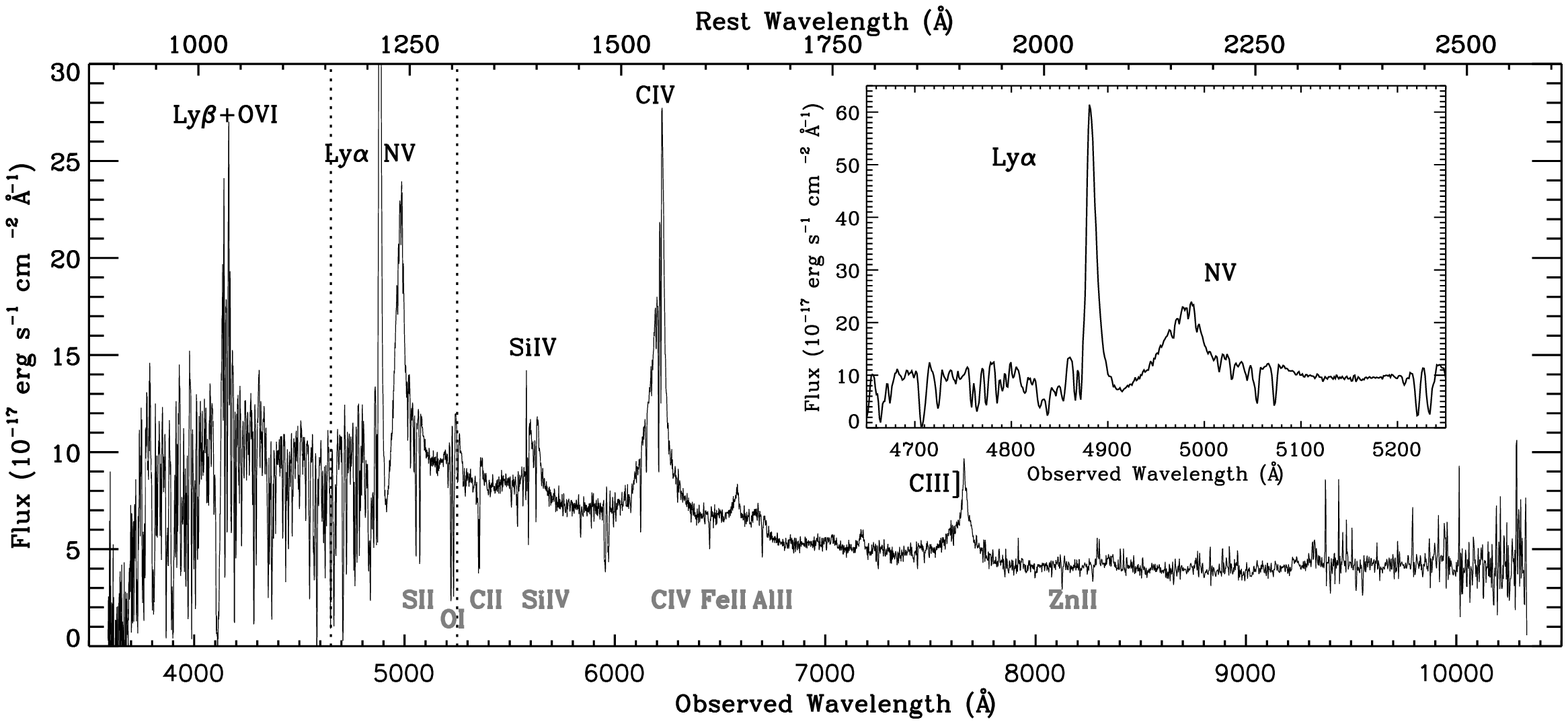}
\caption{\label{fig1} The combined BOSS spectrum of J0952$+$0114. If not stated specifically, rest wavelength in all figures of this paper correspond to the rest-frame of quasar at z$=3.020$. 
The emission lines and the strong absorption lines are labelled
with black and light characters, respectively. The inserted panel zooms in the Ly$\alpha$ and \nv~region indicated by two vertical
dotted lines.}
\end{figure}
\clearpage

\begin{figure}
\epsscale{1.0}
\plotone{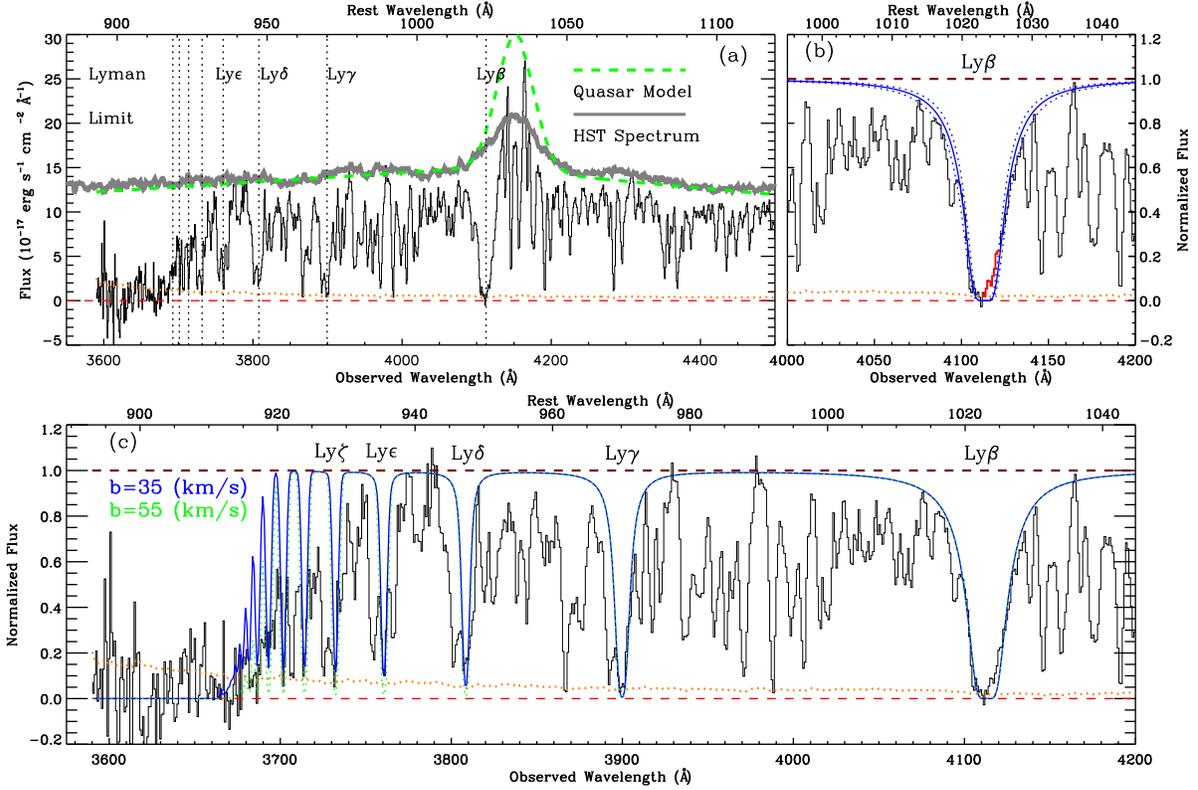}
\caption{\label{fig2}Analysis of Lyman--series absorptions in the spectrum of J0952$+$0114. (a) Normalization of the spectrum: the thick
gray curve is the composite $HST$ quasar spectrum, the dashed green curve is the model of quasar emission constituted with a broken power
law and a Gaussian; (b) Voigt profile fitting to the damped Ly$\beta$ trough: $N_{\rm \hi}=10^{21.8\pm0.2}$~cm$^{-2}$, the residual flux in the line center
is labelled in red; (c) Modelling of the Lyman--series absorptions and the Lyman limit: the solid blue curve is the model with
$N_{\rm \hi}=10^{21.8}$~cm$^{-2}$ and $b=35$~km~s$^{-1}$; the dotted green curve is the model with $N_{\rm \hi}=10^{21.8}$~cm$^{-2}$
and $b=55$~km~s$^{-1}$, having zero fluxes at the line centers of Lyman--series absorptions.
The dotted orange curve is the flux noise in the spectrum.}
\end{figure}
\clearpage

\begin{figure}
\epsscale{1.0}
\plotone{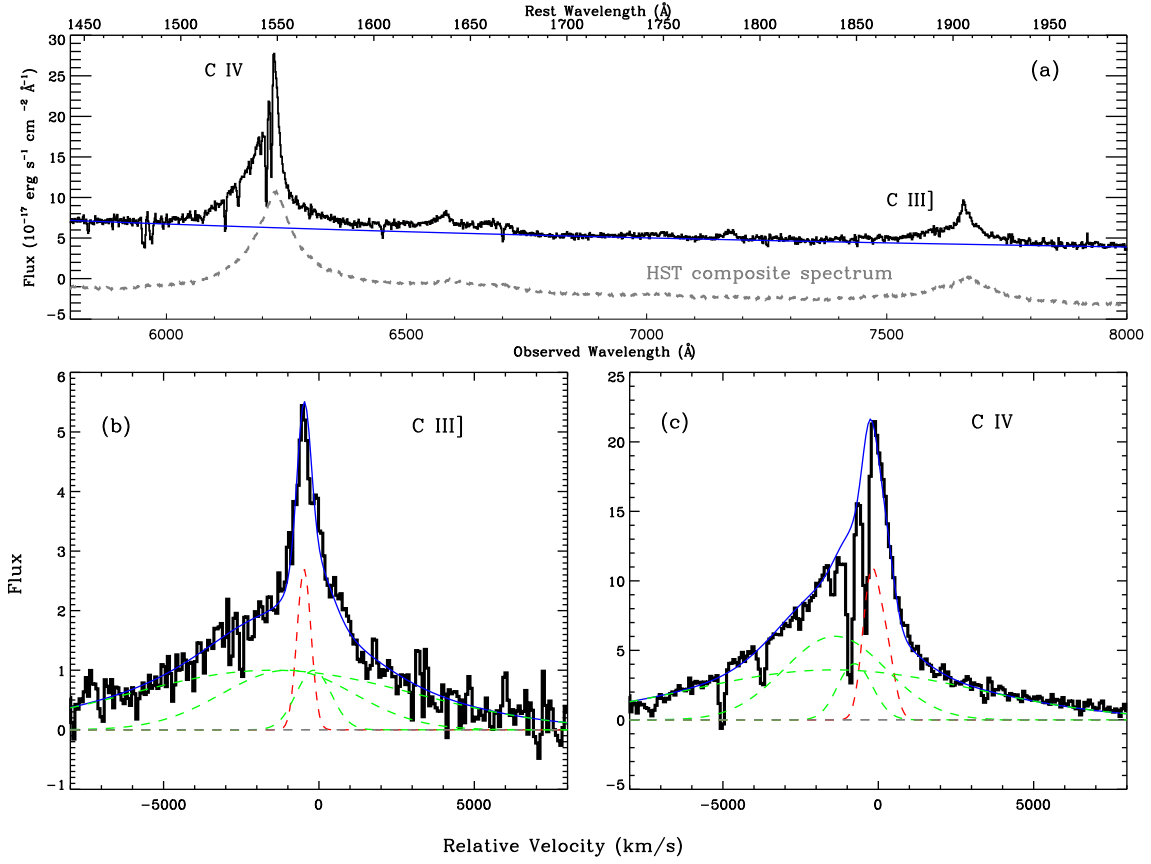}
\caption{\label{fig3}Decomposition of \ciii~and~\civ~emission lines. (a) Comparison of J0952$+$0114's spectrum to the composite $HST$ quasar
spectrum (shifted vertically for illustration purpose). The continuum is fitted with a power law (in blue) and then subtracted from the spectrum;
(b) Decomposition of \ciii~emission line with four Gaussians, the narrow component is in red; (c) Decomposition of \civ~emission line.
} 
\end{figure}
\clearpage

\begin{figure}
\epsscale{1.0}
\plotone{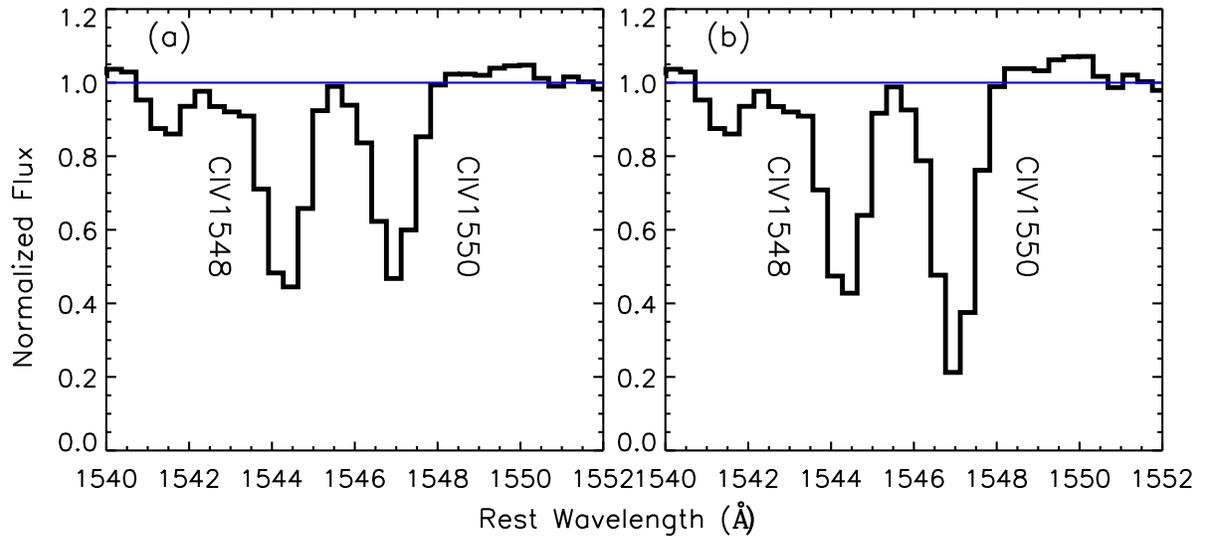}
\caption{\label{fig4}Coverage analysis of the associated \civ~absorption lines. (a) The absorption lines are normalized, assuming the
absorber fully covers the quasar nucleus. (b) The absorption lines are normalized, assuming the absorber does not cover the quasar NELR.
}
\end{figure}
\clearpage

\begin{figure}
\epsscale{1.0}
\plotone{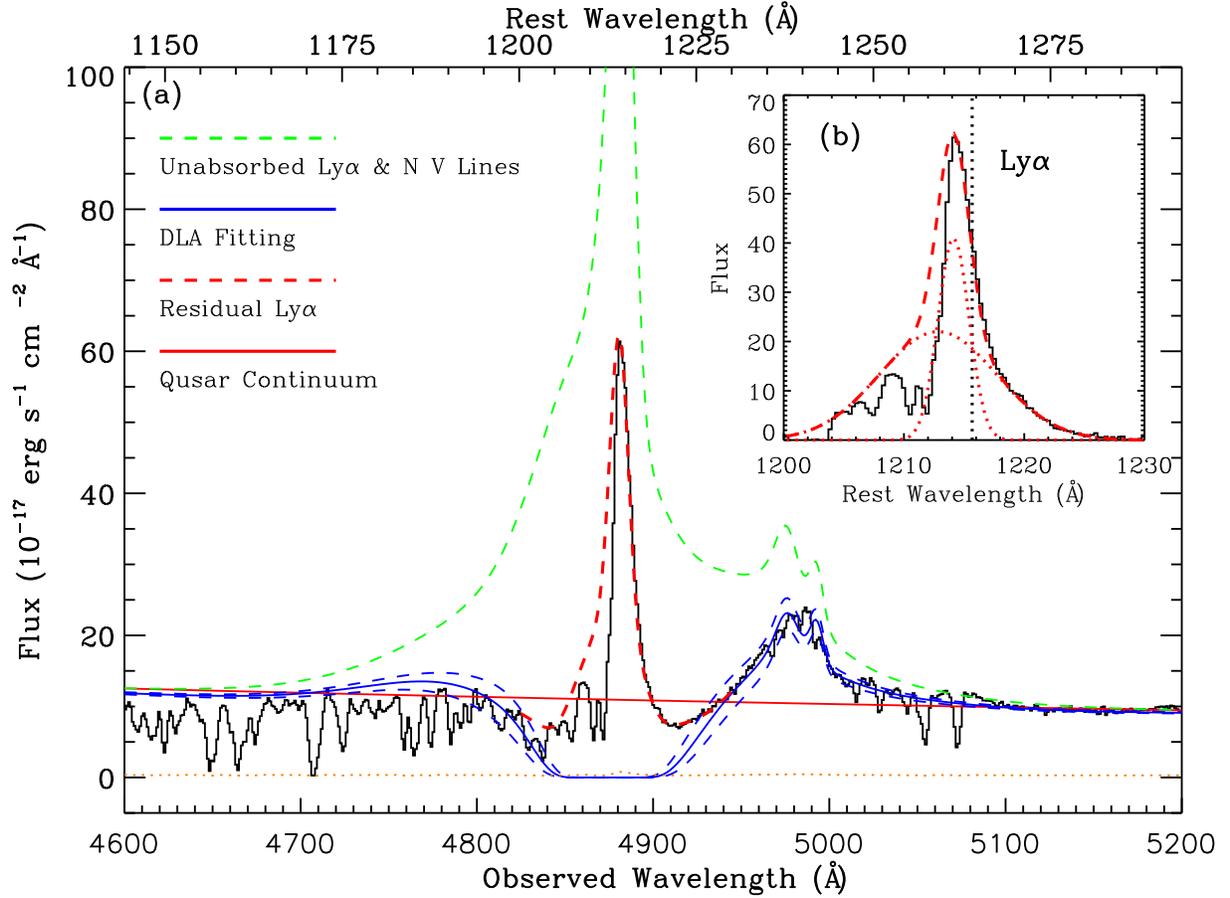}
\caption{\label{fig5} (a) Fitting the DLA profile with the column density of $N_{\rm \hi}=10^{21.8\pm0.2}$~cm$^{-2}$. The unabsorbed quasar
emission consists of a power--law continuum and the broad Ly$\alpha$ and \nv~emission lines, which are modelled with the \civ~emission profile;
(b) The residual Ly$\alpha$ emission is obtained by subtracting the best--fit DLA model from the observed spectrum. It is then fitted with two
Gaussians for correcting the absorptions from Ly$\alpha$ forest.}
\end{figure}
\clearpage

\begin{figure}
\epsscale{1.0}
\plotone{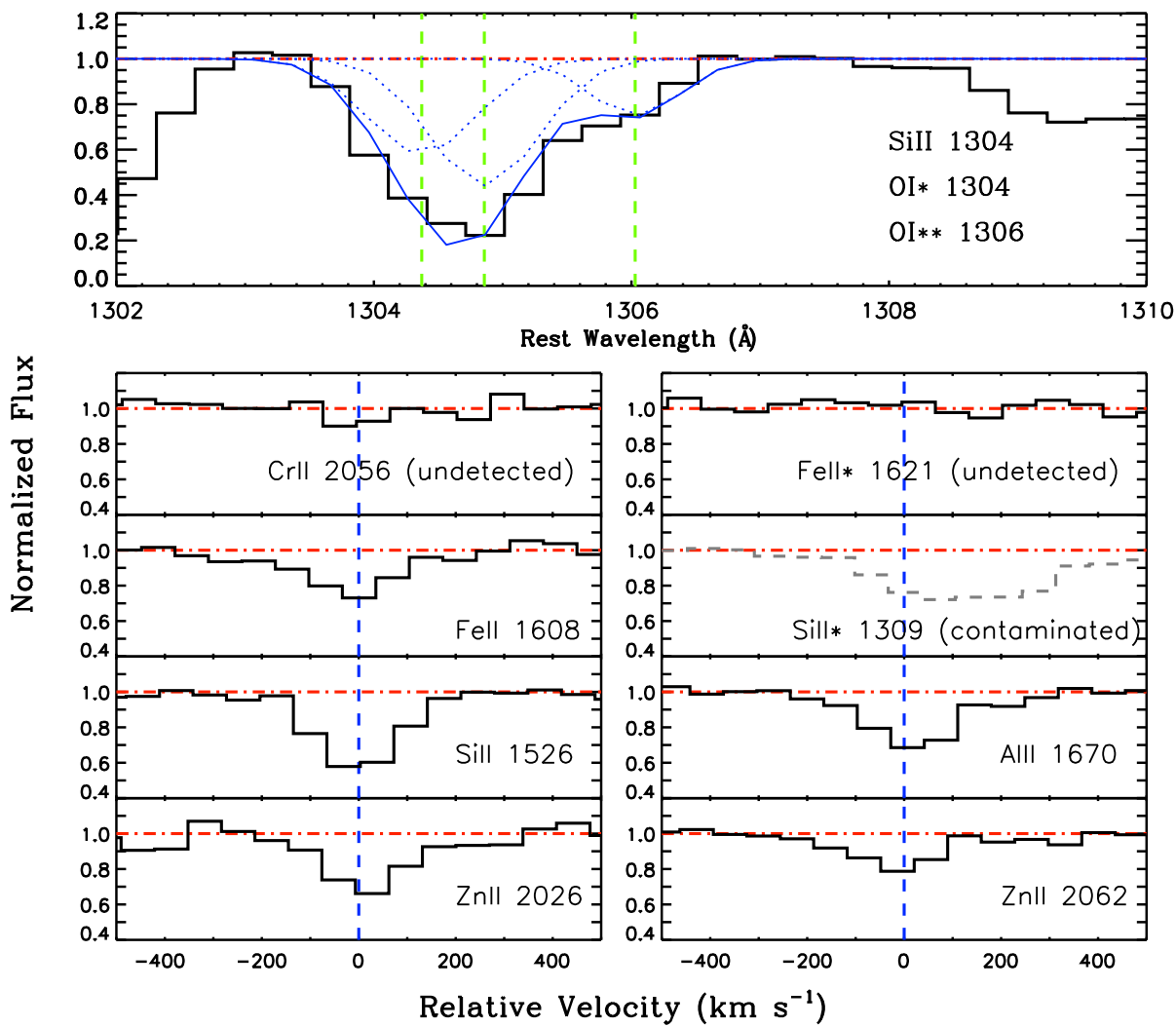}
\caption{\label{fig6}
The associated absorptions lines of interest in J0952$+$0114. The blended absorption profile (solid blue curve) around
$\lambda_{rest}=$1305\AA~is demonstrated at absorber's rest-frame at z$=3.010$. It is decomposed into three Gaussians (thin dotted curves): 
\siii$\lambda$1304, \oi$^*$$\lambda$1304, and \oi$^{**}$$\lambda$1306 with their line centroids marked by green dashed lines from left to right.
In the other panels, zero velocities (blue dashed lines) correspond to  a redshift of z$=3.010$.}
\end{figure}
\clearpage

\begin{figure}
\epsscale{1.0}
\plotone{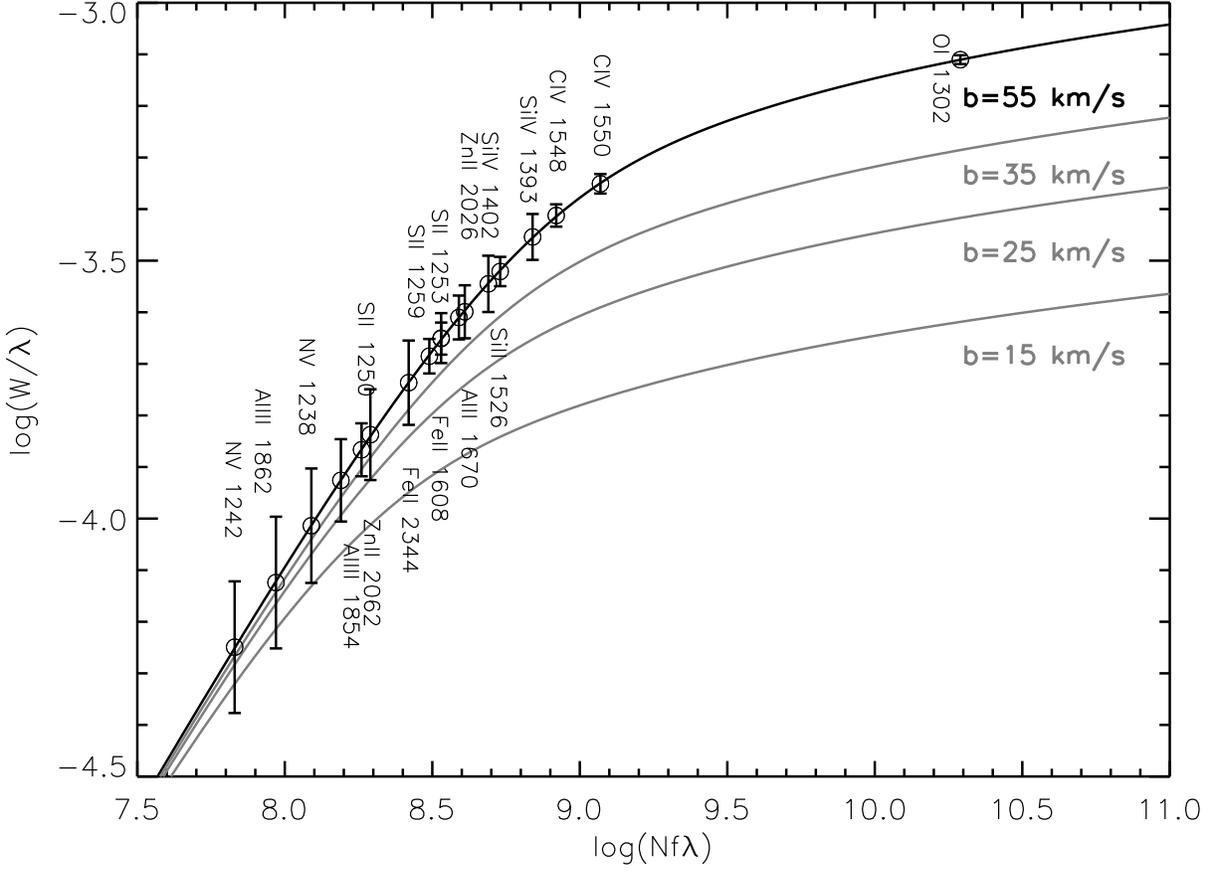}
\caption{\label{fig7}Four curves of growth are created with the Doppler parameters $b=55$ km~s$^{-1}$, $b=35$ km~s$^{-1}$, $b=25$ km~s$^{-1}$
and $b=15$ km~s$^{-1}$ respectively. The associated metal absorption lines are placed onto the curve of $b=55$ km~s$^{-1}$, in order to estimate the lower limits
of their column densities.} 
\end{figure}
\clearpage

\begin{figure}
\epsscale{1.0}
\plotone{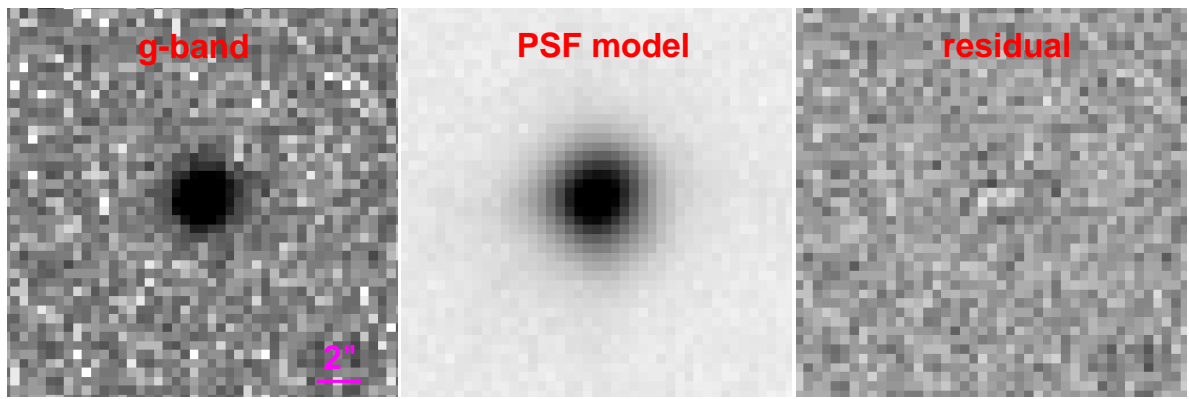}
\caption{\label{fig8}The SDSS image of J0952$+$0114~in $g$ band is in the left panel; the fitted stellar PSF using a 2D Gaussian
with FWHM$=1.3$\arcsec~ is in the middle panel. No significant residual flux is detected in the subtracted image (the right panel).}
\end{figure}
\clearpage

\begin{deluxetable}{llcc}
\tabletypesize{\scriptsize}
\tablecaption{Metal Absorption Lines of the PDLA\label{tbl-1}}
\tablewidth{0pt}
\tablehead{
\colhead{$\lambda$\scriptsize{vacuum}} & \colhead{Ion} & \colhead{EW} & \colhead{Note}\\
\colhead{(\AA)} & \colhead{} & \colhead{(\AA)} &  \colhead{} }
\startdata
1238.821 & N\,\scriptsize{V} & 0.12$\pm$0.03 &  \\
1242.804 & N\,\scriptsize{V} & 0.07$\pm$0.02 & \\
1250.584 & S\,\scriptsize{II} & 0.17$\pm$0.02 & \\
1253.811 & S\,\scriptsize{II} & 0.28$\pm$0.02 & \\
1259.519 & S\,\scriptsize{II} & 0.26$\pm$0.02 & \\
1260.422 & Si\,\scriptsize{II} & 0.75$\pm$0.02 & a \\
1264.737 & Si\,\scriptsize{II}$^*$ & 0.73$\pm$0.02 & a \\
1302.168 & O\,\scriptsize{I} & 1.01$\pm$0.02 &  \\
1304.370 & Si\,\scriptsize{II} & 0.44$\pm$0.03 & b \\
1304.857 & O\,\scriptsize{I}$^*$ & 0.58$\pm$0.03 & b \\
1306.028 & O\,\scriptsize{I}$^{**}$ & 0.26$\pm$0.03 & b \\
1309.275 & Si\,\scriptsize{II}$^*$ & 0.51$\pm$0.03 & c  \\
1334.532 & C\,\scriptsize{II} & 1.22$\pm$0.04 & b \\
1335.707 & C\,\scriptsize{II}$^*$ & 1.22$\pm$0.04 & b \\
1393.755 & Si\,\scriptsize{IV} & 0.49$\pm$0.04 & \\
1402.770 & Si\,\scriptsize{IV} & 0.40$\pm$0.04 & \\
1526.706 & Si\,\scriptsize{II} & 0.46$\pm$0.03 & \\
1533.431 & Si\,\scriptsize{II}$^*$ & 0.33$\pm$0.03 & a \\
1548.195 & C\,\scriptsize{IV} & 0.69$\pm$0.03 & \\
1550.770 & C\,\scriptsize{IV} & 0.60$\pm$0.03 & \\
1608.451 & Fe\,\scriptsize{II} & 0.36$\pm$0.04 & \\
1621.685 & Fe\,\scriptsize{II}$^*$ & $<$0.05 & d \\
1670.787 & Al\,\scriptsize{II} & 0.41$\pm$0.04 & \\
1854.716 & Al\,\scriptsize{III} & 0.22$\pm$0.04 & \\
1862.789 & Al\,\scriptsize{III} & 0.14$\pm$0.04 & \\
2026.136 & Zn\,\scriptsize{II} & 0.51$\pm$0.06 & \\
2056.253 & Cr\,\scriptsize{II} & $<$0.08 & d \\
2062.664 & Zn\,\scriptsize{II} & 0.30$\pm$0.06 & \\
2344.214 & Fe\,\scriptsize{II} & 0.43$\pm$0.08 & \\
\enddata
\tablecomments{Measurements of metal absorption lines in the associated system toward J0952$+$0114.
Error--bars are in 1$\sigma$ level, including statistical errors and systematic errors due to
continuum normalization.
a: blended with intervening absorption lines at lower redshift;
b: blended lines; c: unknown absorption; d: non--detection}
\end{deluxetable}

\end{document}